\documentstyle[12pt]{article}
\newcommand{\R}{\mbox{I \hspace{-0.82em} R}}
\def\bbox{{\,\lower0.9pt\vbox{\hrule \hbox{\vrule height 0.2 cm  

\hskip 0.2 cm 

\vrule  height 0.2 cm}\hrule}\,}}
\begin{document}
\setlength{\unitlength}{1mm}
{\hfill  
{\small DSF-T-44/96, hep-th/9610011}} \\

\vspace{1cm}
\begin{center}
{\Large\bf Heat--kernel Coefficients and Spectra
of the Vector Laplacians on Spherical Domains 
with Conical Singularities}
\end{center}
\bigskip\bigskip\bigskip
\begin{center}
{{\bf Lara De Nardo$^{1}$, Dmitri V.~Fursaev}$^{2,3}$ and 
{\bf Gennaro Miele}$^{1}$}
\end{center}

\bigskip\bigskip

\begin{center}
{\it $^{1}$ Dipartimento di Scienze Fisiche, Universit\`a di Napoli - 
Federico II -, and INFN\\ Sezione di Napoli, Mostra D'Oltremare Pad.  
20, 80125, Napoli, Italy}
\end{center}
\begin{center}
{\it $^{2}$ Theoretical Physics Institute, Department of Physics,  
University of Alberta,\\ Edmonton, Canada T6G 2J1}
\end{center}
\begin{center}
{\it $^{3}$ Joint Institute for Nuclear Research, Bogoliubov
Laboratory of Theoretical Physics,\\ Dubna Moscow Region, Russia}
\end{center}
\vspace*{2cm}
\begin{abstract}
The spherical domains $S^d_\beta$ with conical singularities are a
convenient arena for studying the properties of tensor Laplacians on
arbitrary manifolds with such a kind of singular points. In this paper
the vector Laplacian on $S^d_\beta$ is considered and its spectrum is
calculated exactly for any dimension $d$. This enables one
to find the Schwinger--DeWitt coefficients of this operator by using
the residues of the $\zeta$--function. In particular, the second
coefficient, defining the conformal
anomaly, is explicitly calculated on $S^d_\beta$ and its
generalization to arbitrary manifolds is found. As an application of
this result, the standard renormalization of the one--loop effective
action of gauge fields is demonstrated to be sufficient to remove the
ultraviolet divergences up to the first order in the conical deficit
angle. 
\end{abstract}
\vspace*{1cm}
\noindent
\begin{center}
{\it PACS number(s): 04.60.+n, 12.25.+e, 97.60.Lf, 11.10.Gh}
\end{center}
\vspace*{2cm}
\noindent
e-mail: denardo@na.infn.it; dfursaev@phys.ualberta.ca;  
miele@na.infn.it
\newpage
\baselineskip=.6cm

\section {Motivations and results}
\setcounter{equation}0

Quantum field theory on spaces ${\cal M}_\beta$ with conical
singularities is a research subject which recently many publications
have been devoted to. Conical singularities are a set $\Sigma$ of
points near which ${\cal M}_\beta$ have the structure ${\cal C}_\beta
\times \Sigma$ where ${\cal C}_\beta$ is a cone with an angle $\beta$.
The well--known example of such situation in physics is a space--time
around an  idealized infinitely thin cosmic string \cite{V}. 

Spaces like ${\cal M}_\beta$ appear in the black hole thermodynamics
as well, and for this reason they have been carefully investigated in
the last years. In the Gibbons--Hawking formulation \cite{GH} the free
energy of a black hole is defined in terms of the gravitational action
on the Euclidean black hole instanton. The regularity condition of
such instantons near the Euclidean horizon picks up a period $\beta$
of the imaginary time which is equal to the inverse Hawking
temperature $T_H^{-1}$. On the other hand, the mass $M$ of a black
hole is uniquely related to its Hawking temperature $T_H$ and, hence
the Hamiltonian describing quantum excitations on the black hole
background depends on the temperature as well \cite{Frolov}. However,
the statistical--mechanical definition of the entropy and other
quantities usually requires calculation of the partial derivatives of
the free energy on the temperature. For the black holes it implies
that one has to deal with the temperatures which are different from
the Hawking value and independent on the mass $M$. In this case the
black hole instanton is a space like ${\cal M}_\beta$, and has a
conical singularity on the Euclidean horizon $\Sigma$. The
corresponding approach to the black hole thermodynamics is called the
off--shell approach \cite{CaTe,FFZ}, because singular spaces do not
obey the vacuum Einstein equations. In the Einstein gravity the
off--shell derivation gives the correct Bekenstein--Hawking value for
the black hole entropy \cite{BTZ}. The off--shell method can be also
generalized to the higher--derivative gravity theories \cite{FS95} and
it enables one to get the black hole entropy which agrees with the
other methods. 

The search for the statistical--mechanical explanation of the black
hole entropy \cite{SU} has given rise to the off--shell methods in the
thermodynamics of quantum black holes (see for review \cite{FFZ}). The
methods, which use manifolds with conical singularities, were
formulated and investigated for the static
\cite{SolFur}--\cite{FIS},\cite{FFZ} and for the rotating black holes
\cite{MSa,MSb}. On curved backgrounds the one--loop effective action
$W$ is defined in terms of the trace $K(s)$ of the  heat kernel  for
the Laplace operator $\triangle$ of the corresponding field, by
equation 
\begin{equation}\label{i11}
W=-\frac 12 \log\det \triangle=
-\frac 12 \int_{\delta^2}^{\infty}{ds \over s}K(s)~~~,
\end{equation}
where $\delta^2$ is an ultraviolet cut--off. The important fact is that
the conical singularities change the structure of the ultraviolet
divergences $W_{\mbox{div}}$ of the action (\ref{i11}), which in four
dimensions has the form 
\begin{equation}
W_{\mbox{div}}=-{1 \over 32\pi^2}\left({1 \over 2\delta^4}~A_0 +
{1 \over \delta^2}~A_1-\ln \delta^2~ A_2\right)~~~,
\end{equation}  
because the coefficients $A_n$ of the heat kernel expansion 
\cite{DeWitt}
\begin{equation}\label{i12}
K(s)={1 \over (4\pi s)^2}\left(A_0+s~A_1+s^2~ A_2+...\right)
\end{equation} 
acquire for $n\geq 1$ the additional contributions in the form of
integrals on the singular surface $\Sigma$ \cite{Donnelly},
\cite{CKV}--\cite{Dowker:94PR}. As a consequence of this, the off--shell 
entropy of the quantum fields calculated on ${\cal M}_\beta$ has the
divergence proportional to the area of the horizon $\Sigma$. This
phenomenon has been investigated in detail for scalar fields. In
particular, by following the idea of Ref.\cite{SU}, the important
property that the divergencies of the off--shell black hole entropy are
completely removed under standard renormalization \cite{BirDev} of the
Newton constant and other couplings in the bare gravitational action
was proven \cite{FS96}. 

The analogous properties of the effective actions of higher spin
fields are not well--understood so far. For the vector and Dirac fields
the change of the coefficient $A_1$ in (\ref{i12}) because of conical
singularities has been found in \cite{Kabat,FM96} and the
corresponding ultraviolet divergence is shown to be removed under
renormalization of the Newton constant \cite{Kabat}--\cite{LW}, as in scalar
case. However, as follows from the analysis of \cite{FM96}, the heat
kernel expansion for the spin 3/2 Laplacian and for the Lichnerowicz
operator on ${\cal M}_\beta$ has a feature which makes them different
from the case of the scalar operator. 

This paper studies the vector Laplacians on singular spaces ${\cal
M}_\beta$ and their implications in quantum theory. The way we have
chosen is to consider these operators on $d$--dimensional spherical
domains, denoted further as $S^d_\beta$, and find their spectrum
exactly. Note that in general finding these spectra on curved
manifolds is a quite difficult task, and backgrounds where it can be
done exactly are of a particular interest. 

The space $S^d_\beta$ can be described by the metric 
\begin{equation}\label{i1}
ds^2=a^2\left(\cos^2\chi d\tau^2+d\chi^2 +\sin^2\chi dl^2\right)~~~,
\end{equation}
where $0\leq \tau \leq \beta$, $|\chi|\leq \pi/2$ and $dl^2$ is the
line element on the $(d-2)$--dimensional unit sphere. The points with
coordinates $\chi\simeq \pm \pi/2$ form the sphere $S^{d-2}$ near
which $S^d_\beta$ has the structure ${\cal C}_\beta\times S^{d-2}$.
Outside this domain $S^d_\beta$ coincides with the $d$--dimensional
hypersphere $S^d$ with radius $a$. It should be noted that spaces
$S^d_\beta$ naturally appear in studying the finite-temperature
quantum field theory in static de Sitter space \cite{FM}. 

We show that the Laplace operator $-\nabla_\mu \nabla^\mu$ for the
transverse vector field on $S^d_\beta$ has the following spectrum 
\begin{equation}\label{d1}
\Lambda^{T}_{n,m}(d)={1\over {a}^2}\left[
(n+\gamma m)(n+\gamma m+d-1)-1 \right]~~~, 
\end{equation}
where $\gamma=2\pi/ \beta$, $n=0,1,2,...$ when $m=1,...$, and 
$n=1,2,..$ when $m=0$. The eigen--values with $m=0$ have the
degeneracy 
\begin{equation}
D^{T}_{n,0}(d)={1 \over {n+1}}
\left(
\begin{array}{c}
d+n-3\\n-1\\
\end{array}
\right)
[d^2+(n-4)d+(5-n)]~~~,
\label{deg1}
\end{equation}
while for $m\neq 0$ one has
\begin{equation}
D^{T}_{n,m}(d)=2(d-1)
\left(
\begin{array}{c}
d+n-2\\n\\
\end{array}
\right)~~~.
\label{deg2}
\end{equation}
Then, by using the properties of the $\zeta$--function on $S^d_\beta$
and an assumption about the structure of heat kernel coefficient 
$A_2^{(1)}$ for the vector Hodge--deRham operator $\triangle^{(1)}$ we
find explicitly the addition 
$$
A^{(1)}_{\beta,2}=
{\pi \over 3\gamma}\int_{\Sigma}\left[
(1-\gamma^4){d \over 30}\left(\frac 12R_{ii}-R_{ijij}\right)-
(1-\gamma^2){d \over 6}~R \right.
$$
\begin{equation}\label{i31}
\left.+
(1-\gamma^2)(R-R_{ii})+2(1-\gamma)
(R- 2 R_{ii}+R_{ijij})\right]~~~
\end{equation}
to this coefficient on the arbitrary spaces ${\cal M}_\beta$ with the
conical singularities on $\Sigma$. Here, $R$ is the scalar curvature
and $R_{ii}$, $R_{ijij}$ are invariants constructed of the components
of the Riemann and Ricci tensors normal to $\Sigma$ and computed near
this surface. The knowledge of Eq.(\ref{i31}) enables one to make the
conclusion that the heat-kernel coefficients of the vector
Hodge--deRham operator have the same properties as those of the scalar
Laplacians. In particular, the theorem of Ref.\cite{FS96} about the
renormalization of the entropy, calculated in the conical
singularities method for scalar fields, also holds for vector fields. 

The paper is organized as follows. In Section 2 we start by
considering the scalar Laplacian, and use this case to describe our
strategy and derive some basic formulas. Section 3 is devoted to the
spectrum of the vector Laplacian. The first $A^{(1)}_1$ and second
$A^{(1)}_2$ coefficients in the heat kernel expansion of the vector
Hodge--deRham operator on $S^d_\beta$ are found in Section 4. We use to
this aim the relation between the residues of the $\zeta$--function and
the Schwinger--DeWitt coefficients. The renormalization of the
effective action of the vector fields on ${\cal M}_\beta$ is discussed
in Section 5 after that the results are summarized in Section 6. Some
useful formula and technical details can be found in the Appendices A
and B. 

\section{The method: spectrum of the scalar Laplacian}
\setcounter{equation}0

To find the spectra we will follow the method used in the literature
for the Laplace operators on the hyperspheres $S^d$ (see for instance
\cite{Ordonez}). A $d$--sphere $S^d$ can be embedded into the Euclidean
space $\R^{d+1}$. Let $x^K$, $K=1,..d+1$, be the Cartesian coordinates
in $\R^{d+1}$ and $(r,\tau,\chi,\theta^i)=(r,u^\mu)$, $\mu=1,...d$, be
the spherical ones. The relation between the two coordinate frames
reads 
\begin{equation}\label{s2}
x^{d+1}=r\cos\tau \cos\chi~~,~~x^{d}=r\sin\tau \cos\chi~~,~~x^k=r ~
n^k(\theta)\sin\chi~~,~~k=1,..,d-1,~~~
\end{equation}
where $\theta^i$ (i=1,..d-2) parameterize the unit hypersphere $\sum_k
n_k^2=1$. If the embedding of $S^d$ of radius $a$ in $\R^{d+1}$ is
described by equation $x_Kx^K=a^2$, then the line element on $S^d$ has
the form (\ref{i1}). 

In spherical coordinates the vector field which is in the tangent
space to $S^d$ has components $V^r=0,V^\mu$. For such a field one has
the following relation \cite{Ordonez,LK} 
\begin{equation}
-\nabla^\mu \nabla_\mu V_{\nu}=
\left[-\nabla^K\nabla_K
+\left({\partial \over \partial r}+{{d-1}\over  
r}\right)\left({\partial \over \partial r}-{1 \over r}\right)-
{1 \over {r^2}} \right] V_{\nu}~~~
\label{s1}
\end{equation}
between the Laplacian $-\nabla^\mu \nabla_\mu$ on the hypersphere
$S^d$ with the radius $r$, and the Laplacian $-\nabla^K\nabla_K$ in
$\R^{d+1}$. Therefore, by making use of (\ref{s1}) one can reduce the
problem of finding the spectrum for the vector Laplacian on the
hypersphere to the eigen--problem for the operator $-\nabla^K\nabla_K$
in the flat space, where the latter has a much more easy solution. 

To show that the same method is applicable for the Laplace operator on
the singular $d$--spheres $S^d_\beta$ let us consider the locally flat
space $\R^{d+1}_\beta$. If $\beta<2\pi$, then $\R^{d+1}_\beta$ can be
obtained from $\R^{d+1}$ under identification of the points with the
coordinates $x^K(\tau)$ and $x^K(\tau+\beta)$. When $\beta > 2\pi$,
one has at first to replace $\R^{d+1}$ with the space $\R^{d+1}_\infty$, 
which has an infinite range for the polar coordinate $\tau$ defined in
(\ref{s2}), and represents an analog of the
Riemann sheet. Then $\R^{d+1}_\beta$ appears after the identification
of the points $x^K(\tau)$ with $x^K(\tau+\beta)$ of $\R^{d+1}_\infty$. 

Although $\R^{d+1}_\beta$ is a locally flat space, it has the conical
singularities on the $(d-1)$--dimensional hyperplane $x^{d+1}=x^d=0$
($\chi=\pm \pi/2$). The singular sphere $S^d_\beta$ with the metric
(\ref{i1}) can be embedded into $\R^{d+1}_\beta$. This embedding is
described by the same equation $x_Kx^K=a^2$ as in the previous case,
provided the relation between Cartesian coordinates $x^K$ and
spherical ones $(r,\tau,\chi,\theta^i)$ is left unchanged. As it is
easy to understand, the relation between the vector Laplacian in
$\R^{d+1}_\beta$ and on $S^d_\beta$ does not change and coincides with
the Eq.(\ref{s1}).  Therefore for the spectral problem it is
sufficient to investigate the operator $-\nabla^K \nabla_K$ in this
locally flat space. 

\bigskip\bigskip

At first it is worth applying the above technique to find the spectrum
for the scalar Laplacian on $S^d_{\beta}$. The results will be useful
then in studying the vector case. Let us consider to this aim the
scalar field on $S^d_\beta$, whose wave functions $\phi$ belong to the
Hilbert space $L^2(S^d_\beta)$ and are periodic when $\tau$ is
increased by $\beta$ 
\begin{equation}\label{per1}
\phi(\tau+\beta,u)=\phi(\tau,u)~~~.
\end{equation}
We show below that for $\beta \leq 2\pi$ the eigen--functions
$\phi_{n,m}^{\pm}$ of this operator can be represented as homogeneous
polynomials in the Cartesian coordinates 
\begin{equation}\label{egv}
\phi_{n,m}^{\pm}=r^{-(n+\gamma m)}\sum_{p=0}^{[n/2]} 
C^{\pm m}_{i_1....i_{n-2p}}x^{i_1}...
x^{i_{n-2p}}(x^{+}x^{-})^{p}
(x^{\pm})^{\gamma m}~~~,
\end{equation}
where $[n/2]$ is the integer part of $n/2$, and $C^{\pm
m}_{i_1,...i_{n-2p}}$ are symmetric tensors. Note that the coordinates
$x^{\pm}$, defined as 
\begin{equation}\label{s4}
x^{\pm}\equiv{1 \over \sqrt{2}}\left(x^{d+1}\pm i x^{d}\right)~~~,
\end{equation}
transform as $x^{\pm}(\tau+\beta)=e^{\pm i\beta}x^{\pm}(\tau)$.
The corresponding eigen--values for both $\phi_{n,m}^{+}$ and
$\phi_{n,m}^{-}$ read 
\begin{equation}\label{s5}
\Lambda^{(0)}_{n,m}={1 \over {a^2}}~(n+\gamma m)
(n+\gamma m+d-1)~~~,~~~n,m=0,1,2,...~~~
\end{equation}
and have the following degeneracy
\begin{equation}\label{s10}
D^{(0)}_{n,m=0}(d) = \left(
\begin{array}{c}
n+d-2\\n\\
\end{array}
\right)
\end{equation}
and $D^{(0)}_{n,m\neq 0}(d)=2D^{(0)}_{n,m=0}(d)$.

\bigskip\bigskip

To prove this result let us consider a polynomial
of an arbitrary form
\begin{equation}\label{egv2}
\phi_{n,\omega,\rho}=r^{-(n+\omega+\rho)}\sum_{q=0}^n 
C^{\omega,\rho}_{i_1....i_{n-q}}x^{i_1}...
x^{i_{n-q}}(x^{+}x^{-})^{\frac 12 q}
(x^{+})^{\omega}(x^{-})^{\rho}~~~,
\end{equation}
where $\omega$ and $\rho$ are some constants which do not depend on
$n$. This polynomial obeys the generalized boundary condition 
\begin{equation}\label{per2}
\phi_{n,\omega,\rho}(\tau+\beta,u)=
e^{i(\omega-\rho)\beta}\phi_{n,\omega,\rho}(\tau,u)~~~,
\end{equation}
and does not depend on $r$,
\begin{equation}\label{r}
{\partial \over \partial r}
\phi_{n,\omega,\rho}=0~~~.
\end{equation}
As it is easy to see the functions (\ref{egv2}) belong to
$L^2(S^d_\beta)$ if 
\begin{equation}\label{intcond}
\rho +\omega > -1~~~.
\end{equation}
A direct computation shows that 
$\phi_{n,\omega,\rho}$
are eigen--functions of the operator $-\nabla^K\nabla_K$
\begin{equation}\label{egval}
-\nabla^K\nabla_K \phi_{n,\omega,\rho}=
{1 \over r^2}(n+\omega +\rho)
(n+\omega +\rho+d-1)
\phi_{n,\omega,\rho}
\end{equation}
provided the following conditions:
\begin{equation}\label{cond1}
\omega\rho~ C^{\omega,\rho}_{i_1,....i_n}=0~~~,
\end{equation}
\begin{equation}\label{cond2}
(2\omega+1)(2\rho+1)C^{\omega,\rho}_{i_1,....i_{n-1}}=0~~~,
\end{equation}
\begin{equation}\label{cond3}
\frac 12(2\omega+q)(2\rho+q)C^{\omega,\rho}_{i_1,....i_{n-q}}
+(n-q+2)(n-q+1)C^{\omega,\rho}_{j,j,i_1,...i_{n-q}}=0~~~
\end{equation}
are imposed. It is assumed in Eq.(\ref{cond3}) that $q\geq 2$ and the
summation is taken over the repeating indexes. If the coefficient
$C^{\omega,\rho}_{i_1,...i_n}\neq 0$, it defines all other
coefficients $C^{\omega,\rho}_{i_1,..i_{n-q}}$ with $q=2p$. In this
case equation (\ref{cond1}) can be satisfied when $\omega=0$ or
$\rho=0$. Then condition (\ref{cond2}) holds when all other
coefficients $C^{\omega,\rho}_{i_1,..i_{n-q}}$ with odd $q=2p+1$ are
zero. 

\bigskip

For the field (\ref{per1}) the additional periodicity condition has to
be imposed. Then, the form (\ref{egv}) of the eigen--vectors
$\phi_{n,m}^{\pm}$ follows from Eq.(\ref{egv2}) where the coefficients
for odd $q$'s have to be omitted. The functions with $\omega=0$ and
$\rho=m\gamma$ correspond to the eigen--modes $\phi_{n,m}^{-}$, while
the functions with $\rho=0$ and $\omega=m\gamma$ correspond to
$\phi_{n,m}^{+}$. If $\beta \leq 2\pi$ ($\gamma \geq 1$), then
$m=0,1,2,...$. If $\beta> 2\pi$, there may be a finite number of
eigen--modes, belonging to $L^2(S^d_\beta)$, with negative $m$, which
are, however, singular on the hypersurface $x^{+}= x^{-}=0$. In this
case an additional analysis is required. For this reason our approach
is to find the results for $\beta \leq 2 \pi$ and then to extend them
for any positive $\beta$ by means of the analytical continuation. 

Finally by taking into account Eqs.(\ref{r}), (\ref{egval}) and the
fact that the scalar operators $-\nabla^K\nabla_K$ and
$-\nabla^\mu\nabla_\mu$ coincide, one gets the eigen--values
(\ref{s5}) on the sphere $S^d_\beta$ with radius $r=a$. 

To compute the degeneracy, let us note that in (\ref{egv}) all the
coefficients $C^{\pm m}_{i_1,...i_{n-2p}}$ are determined by
(\ref{cond3}) in terms of the $(d-2)$--dimensional rank--$n$ symmetric
tensor $C^{\pm m}_{i_1,...,i_n}$. The number of components 
\begin{equation}
\left(
\begin{array}{c}
n+d-2\\n\\
\end{array}
\right)~~~
\label{6}
\end{equation}
of $C^{\pm m}_{i_1,...,i_n}$ gives the number of independent
eigen--modes $\phi^{\pm}_{n,m}$. The degeneracy $D^{(0)}_{n,m=0}$
coincides with the number of modes with $m=0$. The degeneracy
$D^{(0)}_{n,m\neq 0}= 2D^{(0)}_{n,m=0}$ is the total number of
$\phi^{+}_{n,m}$ and $\phi^{-}_{n,m}$ because they correspond to the
equal eigen--values. 

\bigskip

It is instructive to see how the well--known spectrum of the Laplace  
operator on $S^d$ follows from our results. If $\beta=2\pi$ the 
eigen--values (\ref{s5}) depend on the number $l=n+m$ only
\begin{equation}\label{s12}
\Lambda^{(0)}_l(d)=\Lambda^{(0)}_{l,l-n}(d)= {{l~(l+d-1)}\over a^2}~~~,
~~~\beta=2\pi~~ .
\end{equation}
The total degeneracy $D^{(0)}_l(d)$ corresponding to a given $l$ is  
the sum
\begin{equation}\label{s13}
D^{(0)}_l(d)=D^{(0)}_{l,m=0}(d)+2\sum^{l}_{m=1}
D^{(0)}_{l-m,0}(d)={(d+l-2)!\over{l!(d-1)!}}(2l+d-1)~~~,
\end{equation}
which can be checked with the help of the relation
\begin{equation}\label{summa}
\sum^m_{k=0} \left(
\begin{array}{c}
h+k\\h\\
\end{array}
\right)=\left(
\begin{array}{c}
h+m+1\\h+1\\
\end{array}
\right)~~~.
\end{equation}
The eigen--values (\ref{s12}) and degeneracies (\ref{s13}) coincide
with the known results \cite{Ordonez}. The spectrum of the scalar
operator on $S^d_\beta$ in the form (\ref{s5}), (\ref{s10}) was first
established in \cite{FM} for the four--dimensional case. 

\section{Spectrum of the vector Laplacian}
\setcounter{equation}0

\subsection{Eigen--values}
We denote the Cartesian components of a vector with $V^K$. For a
vector from the tangent space to $S^d_\beta$, whose components are
defined with respect to the spherical coordinates, the periodicity
condition takes the simple form 
\begin{equation}\label{v2}
V^\mu(\tau+\beta)=V^\mu(\tau)~~~,
\end{equation}
while in the Cartesian frame it looks as
\begin{equation}\label{Vv}
V^L(\tau+\beta)=O_K^L(\beta)~V^K(\tau)~~~.
\end{equation}
The matrix $O_K^L(\beta)=\partial x^L(\tau+\beta)/\partial x^K(\tau)$
appears because the Cartesian basis $\partial /\partial x^L$ does not
perform the complete rotation when $\tau$ is increased by $\beta$. 

On $d$--spheres the eigen--value problem for the vector Laplacian is
reduced to the separate problems for the transversal and longitudinal
components of the vector field $V_{\mu}$. The spectrum for the
longitudinal part on $S^d_\beta$ is completely determined by the
scalar spectrum. This follows from the relation 
\begin{equation}\label{v1}
-\left(\nabla^\mu \nabla_\mu\delta_\rho^\nu  
-R_\rho^\nu\right)\nabla_\nu \phi = 
-\nabla_\rho \left(\nabla^\mu  
\nabla_\mu \phi\right)~~~,
\end{equation}
where $\phi$ is a scalar field and
$R^\rho_\nu=a^{-2}(d-1)\delta^\rho_\nu$ is the Ricci tensor computed
in the regular domain of $S^d_\beta$. For this reason we will consider
further only transversal vectors. 

\bigskip

At first let us find out the general form of the eigen--vectors $V^L$
for the operator $-\nabla_K\nabla^K$ restricted to $S^d_\beta$. The
results of Section 2 can be helpful for this aim, if we consider the
basis connected with the coordinates $(x^{+},x^{-},x^k)$,
$k=1,..,d-1$. Then instead of the components $V^{d}$ and $V^{d+1}$ one
has the components\footnote{Note that $V_{\mp}=V^{\pm}$.} 
\begin{equation}\label{v5}
V^{\pm}={1 \over \sqrt{2}}\left(V^{d}\pm iV^{d+1}\right)~~~,
\end{equation}
which transform under rotations (\ref{Vv}) as
\begin{equation}\label{v6}
V^{\pm}(\tau+\beta)=e^{\pm i\beta}V^{\pm}(\tau)~~~.
\end{equation}

Obviously, the components $V^L$ can be treated just as a set of scalar
fields. Thus, it follows from Eqs.(\ref{egv2}),(\ref{egval}) and
conditions (\ref{cond1}),(\ref{cond2}) that vector eigen--modes can be
written in the following form 
\begin{equation}\label{v4}
\left(
\begin{array}{c}
(V^+)_{n,m}\\ \\ (V^k)_{n,m}\\ \\(V^{-})_{n,m}\\
\end{array}
\right)=
r^{-(n+\gamma m)}\left(
\begin{array}{c}
(x^+)^{\gamma m +1} \sum_{p=0}^{[{n-1\over 2}]}(B^+)^m_{i_1,...,i_{n-2 p-1}}
x^{i_1}\cdot\cdot\cdot x^{i_{n-2 p-1}}~(x^{+}x^-)^{p}\\
\\
(x^+)^{\gamma m}\sum_{p=0}^{[{n\over 2}]}(B^k)^m_{i_1,...,i_{n-2 p}}
x^{i_1}\cdot\cdot\cdot x^{i_{n-2 p}}~(x^{+}x^-)^{p}\\
\\
(x^+)^{\gamma m-1} \sum_{p=0}^{[{n+1\over 2}]}(B^-)^m_{i_1,...,i_{n-2 p+1}}
x^{i_1}\cdot\cdot\cdot x^{i_{n-2 p+1}}~(x^{+}x^-)^{p}\\
\end{array}
\right)~~~,
\label{vv1}
\end{equation}
where the quantities $(V^+)_{0,m}$ are assumed to vanish. The range of
$n$ and $m$ is $n=0,1,2,...$ when $m=1,2,...$, and $n=1,2,...$ when
$m=0$. It is easy to check that for $n=m=0$ there are no transverse
modes. For simplicity we presented only one set of the eigen--modes.
Another set can be obtained from (\ref{v4}) under interchange of $x^+$
with $x^-$. According to our notation, $(B^{K})^m_{i_1,...,i_n}$
denote symmetric tensors, or constants, if the index is absent. We
also impose the additional restriction at $m=0$ 
\begin{equation}\label{cond4}
(B^-)^{0}_{i_1,...,i_{n+1}}=0~~~.
\end{equation}
Then all the components in (\ref{v4}) are in the Hilbert
space $L^2(S^d_\beta)$.

The vector (\ref{v4}) has the required periodicity and it is
constructed in such a way that all its components have the same
eigen--value for given numbers $n$ and $m$. It can be verified with
the help of (\ref{egval}) that 
\begin{equation}\label{egval3}
-\nabla^K\nabla_K (V^L)_{n,m} = {1 \over a^2}(n+\gamma m )
(n+\gamma m +d-1) (V^L)_{n,m}~~~,
\end{equation}
provided the conditions
\begin{eqnarray}
2 p (p + \gamma m - 1) 
(B^{-})^m_{i_1,...,i_{n-2p+1}} +(n-2 p +2) (n-2 p +3)
(B^{-})^m_{h,h,i_1,...,i_{n-2p+1}} &=&0~~~\nonumber\\ 
\mbox{for}~~~1 \leq p
\leq\left[{n+1 \over 2} \right]~~~,&~&\label{v13}\\  
2 p (p + \gamma m) 
(B^{k})^m_{i_1,...,i_{n-2p}} +(n-2 p +1) (n-2 p +2)
(B^{k})^m_{h,h,i_1,...,i_{n-2p}} &=&0~~~\nonumber\\
\mbox{for}~~~1 \leq p     
\leq\left[{n \over 2} \right]~~~,&~&\label{v14}\\   
2 p (p + \gamma m + 1) 
(B^{+})^m_{i_1,...,i_{n-2p-1}} +(n-2 p ) (n-2 p +1)
(B^{+})^m_{h,h,i_1,...,i_{n-2p-1}} &=&0~~~\nonumber\\
\mbox{for}~~~1 \leq p
\leq\left[{n-1 \over 2} \right]~~~,&~&\label{v15} 
\end{eqnarray}
which follow from Eq.(\ref{cond3}), are satisfied.

Note now that, as in the scalar case, the Cartesian components $V^L$
obey Eq.(\ref{r}) 
\begin{equation}\label{r2}
{\partial V^L \over \partial r}=0~~~.
\end{equation}
The same condition for the covariant components in the spherical
coordinates used in Eq.(\ref{v1}) reads
\begin{equation}\label{r3}
{\partial V_\mu \over \partial r}=\frac 1r V_\mu~~~,
\end{equation}
because the matrix which defines the relation
$$
V_\mu ={\partial x^L \over \partial u^\mu} 
V_L
$$ 
between two systems, scales as $r$. Eq.(\ref{r3}) does not change the
action of the operator $-\nabla^K\nabla_K$ on $(V_\mu)_{n,m}$, 
\begin{equation}\label{egval5}
-\nabla^K\nabla_K (V_\mu)_{n,m} = 
{1 \over a^2}
(n+\gamma m)
(n+\gamma m +d-1)(V_\mu)_{n,m}~~~.
\end{equation}
Therefore according to Eqs.(\ref{v1}) and (\ref{r3}) we
have the formula
\begin{equation}\label{egval4}
-\nabla^\nu\nabla_\nu (V_\mu)_{n,m} 
={1 \over a^2}\left[
(n+\gamma m)
(n+\gamma m +d-1)-1\right](V_\mu)_{n,m} =
\Lambda^{T}_{n,m}(V_\mu)_{n,m}~~~,
\end{equation}
which gives the eigen--values $\Lambda^{T}_{n,m} $ in Eq.(\ref{d1}).

\subsection{Degeneracies}
We must now ensure that the vector field (\ref{v4}) is transverse and
it is in the tangent space to $S^d_\beta$. These conditions read 
\begin{equation}\label{v9}
x^K V_K=x^i V^i+x^{+} V^{-}+x^{-} V^{+}=0~~~,
\end{equation}
\begin{equation}\label{v8}
\nabla_K V^K=\nabla_j V^j+\nabla_{+} V^{+}+\nabla_{-} V^{-}=0~~~,
\end{equation}
where in the Cartesian basis $\nabla_K=\partial /\partial x^K$ and
$\nabla_{\pm}=\partial /\partial x^{\pm}$. The first condition
(\ref{v9}), after substitution of expression (\ref{v4}) takes the form
$$
\sum_{p=1}^{[{n+1\over 2}]}(B^+)^m_{i_1,...,i_{n-2 p+1}}
x^{i_1}\cdot\cdot\cdot x^{i_{n-2 p+1}}~(x^{+}x^-)^{p}+
\sum_{p=0}^{[{n\over 2}]}(B^k)^m_{i_1,...,i_{n-2 p}}
x^{i_1}\cdot\cdot\cdot x^{i_{n-2 p}} x^k~(x^{+}x^-)^{p}
$$
\begin{equation}
+\sum_{p=0}^{[{n+1\over 2}]}(B^-)^m_{i_1,...,i_{n-2 p+1}}
x^{i_1}\cdot\cdot\cdot x^{i_{n-2 p+1}}~(x^{+}x^-)^{p}=0~~
\label{vv2}
\end{equation}
which is equivalent to a number of constraints
\begin{equation}\label{cond21}
(n-2p+1)\left[(B^+)^m_{k,i_1,...,i_{n-2 p}} +
(B^-)^m_{k,i_1,...,i_{n-2 p}}\right] + (B_{\{k})^m_{i_1,...,i_{n-2 p}
\} }  =  0~~~, 
\end{equation}
where $1\leq p \leq [(n-1)/2]$ and $n>1$, and
\begin{equation}\label{cond22}
(B_{\{k})^m_{i_1,...,i_{n}\}}+
 (n+1)  (B^-)^m_{k,i_1,...,i_{n}}  =  0~~~,
\end{equation}
\begin{equation}\label{cond23}
(B^+)^m + (B^-)^m  =  0 ~~~,~~~\mbox{for}~~n~~\mbox{odd}~~~.
\end{equation}
Here $ (B_{\{k})^m_{i_1,...,i_{n-2p}\}}$ is the expression which is
completely symmetric in the all lower indexes, namely, 
\begin{equation}
(B_{\{k})^m_{i_1,...,i_{n-2p}\}}\equiv
(B_k)^m_{i_1,...,i_{n-2 p}} +
\sum_{l=1}^{n-2p} (B_{i_l})^m_{i_1,...,i_{l-1},k,i_{l+1}...,i_{n-2p}}~~~,
\end{equation}
$$
(B_k)^m_{i_1,...,i_{n-2 p}}=(B^k)^m_{i_1,...,i_{n-2 p}}~~~.
$$
Finally, if one uses (\ref{v9}), then transversality condition (\ref{v8}) 
results in equations
\begin{equation}\label{cond24}
(\gamma m + p +1)(B^+)^m_{i_1,...,i_{n-2 p-1}} + 
(p + 1) (B^-)^m_{i_1,...,i_{n-2 p-1}} + (n - 2 p)
(B_h)^m_{h,i_1,...,i_{n-2 p-1}}  =  0~~~,
\end{equation}
where  $n \geq 1$ and $0\leq p \leq [(n -1)/2]$.
As follows from restriction (\ref{cond4}), in counting the degeneracies
we should consider the cases $m\neq 0$ and $m=0$ separately.

\bigskip

\noindent
{\it a) $m\neq 0$} 

\bigskip

\noindent
For these values of $m$, Eqs.(\ref{v13})--(\ref{v15}) enable one to
express all the coefficients $(B^K)^m_{i_1,....i_k}$ in terms of the
coefficients $(B^K)^m_{i_1,...,i_{k+2}}$. Therefore the number of
independent constants in (\ref{v4}) is completely determined by the
number of independent quantities $(B^-)^m_{i_1,...,i_{n+1}}$,
$(B^k)^m_{i_1,...,i_n}$, and $(B^+)^m_{i_1,...,i_{n-1}}$ . However,
$(B^+)^m_{i_1,...,i_{n-1}}$ are combinations of
$(B^-)^m_{i_1,...,i_{n+1}}$ and $(B^k)^m_{i_1,...,i_n}$ because of
Eq.(\ref{cond21}) with $p=1$.  On the other hand,
$(B^-)^m_{i_1,...,i_{n+1}}$ are given in terms of
$(B^k)^m_{i_1,...,i_n}$ by Eq.(\ref{cond22}). The explicit resolution
of the constraints is represented in Appendix A by formulas
(\ref{v16})--(\ref{v18}). Let us note that in this way we get the
coefficients which by the construction obey relations
(\ref{v13})--(\ref{v15}). What is more non--trivial is that also
Eqs.(\ref{cond21}) and (\ref{cond24}) hold automatically on
(\ref{v16})--(\ref{v18}) for {\it any} number $p$. The same is true
for Eq.(\ref{cond23}). This can be checked by the direct substitution.

Thus the number of independent constants in (\ref{v4}) coincides with
the number of components of the  $(d-1)$ symmetric rank--$n$ tensors
$(B^k)^m_{i_1,...,i_n}$. The degeneracy 
\begin{equation}
\label{v19}
D^{T}_{n,m\neq 0}(d)=2 (d-1) \left(
\begin{array}{c}
n+d-2\\n\\
\end{array}
\right)
\end{equation}
is twice of this number because apart (\ref{v4}) one can construct (by
interchanging $x^+$ and $x^-$ in (\ref{v4})) the analogous set of
vectors with the same eigen--values. 

\bigskip

\noindent
{\it b) $m=0$} 

\bigskip

\noindent 
In this case, as follows from Eqs.(\ref{v13})--(\ref{v15}) and
(\ref{cond4}), the number of independent constants in (\ref{v4}) is
completely determined by the number of independent quantities
$(B^-)^0_{i_1,...,i_{n-1}}$, $(B^k)^0_{i_1,...,i_n}$, and
$(B^+)^0_{i_1,...,i_{n-1}}$. The tensors $(B^+)^0_{i_1,...,i_{n-1}}$
are combinations of $(B^-)^0_{i_1,...,i_{n-1}}$ and
$(B^k)^0_{i_1,...,i_n}$ because of Eq.(\ref{cond21}) with $p=1$.
However, $(B^-)^0$--tensors are not related now to
$(B^k)^0_{i_1,...,i_n}$ via (\ref{cond22}), as it was in case $m \neq
0$, and they are determined in terms of $(B^-)^0_{i_1,..,i_{n-1}}$.
Moreover from conditions (\ref{cond4}) and (\ref{cond22}) one gets the
additional constraint 
\begin{equation}\label{cond25}
(B_{\{k})^0_{i_1,...,i_n\}}=0~~~.
\end{equation}
In summary, all the coefficients for $m=0$ can be represented in terms
of constants $(B^-)^0_{i_1,..,i_{n-1}}$ and $(B^k)^0_{i_1,...,i_n}$
with restriction (\ref{cond25}). The explicit form of the solution is
given in Appendix A by Eqs.(\ref{v20})--(\ref{v23}). Again by the
direct substitution of these solutions one can check that they satisfy
Eq.(\ref{cond23}) and constraints (\ref{cond21}), (\ref{cond24}) for
any $p$. Thus the degeneracy of the eigen--value with $m=0$ 
\begin{eqnarray}
D^{T}_{n,0}(d)&=&\left(
\begin{array}{c}
n-1+d-2\\d-2\\
\end{array}
\right)
+(d-1)\left(
\begin{array}{c}
n+d-2\\d-2\\
\end{array}
\right)-  
\left(
\begin{array}{c}
n+1+d-2\\d-2\\
\end{array}
\right)\nonumber\\
&=&{1 \over {n+1}}
\left(
\begin{array}{c}
d+n-3\\n-1\\
\end{array}
\right)
[d^2+(n-4)d+(5-n)]~~~.
\label{v27}
\end{eqnarray} 
is given by the number of symmetric tensors
$(B^-)^0_{i_1,...,i_{n-1}}$ and $(B^k)^0_{i_1,...,i_n}$ minus the
number of conditions (\ref{cond25}). 

\bigskip

The eigen--values (\ref{egval4}) and degeneracies (\ref{v19}),
(\ref{v27}) obtained on singular spheres $S^d_\beta$ reproduce in the
limit $\beta=2\pi$ ($\gamma=1$) the results of \cite{Ordonez} for
$d$--spheres. Indeed, if $\gamma=1$ the eigen--values (\ref{egval4})
can be expressed in terms of a single number $l=n+m$, 
\begin{equation}
\Lambda^{T}_{l}(d)=
\Lambda^{T}_{n,l-n}(d)
={1 \over a^2}\left(l(l +d-1)-1\right)~~~,~~~l=1,2,...~~~,
\end{equation}  
and coincide with the eigen--values for the transverse vectors in
\cite{Ordonez}. Analogously, the total degeneracy is obtained from the
equation 
\begin{equation}
D^{T}_l(d)=D^{T}_{l,m=0}(d)+2\sum^{l}_{m=1}
D^{T}_{l-m,0}(d)={l(l+d-1)(2l+d-1)(l+d-3)!
\over (d-2)!(l+1)!}~~~,
\end{equation}
with the help of summation formula (\ref{summa}) and it coincides with
the result \cite{Ordonez}. 

\section{Heat--coefficients of vector Hodge--deRham operator}
\setcounter{equation}0

\subsection{Relation to $\zeta$--function}
We now consider $\zeta$--function and heat--coefficients
for the vector Hodge--deRham operator $\triangle^{(1)}$ which 
is defined as
\begin{equation}\label{h1}
\triangle^{(1)} V^{\nu}=
-\left(\nabla^\mu \nabla_\mu\delta_\rho^\nu  
-R_\rho^\nu\right) V^{\rho}~~~,
\end{equation}
where $R^\rho_\nu=a^{-2}(d-1)\delta^\rho_\nu$ is the Ricci tensor
computed in the regular domain of $S^d_\beta$. As we mentioned before
the transversal and longitudinal parts of the vector field have to be
studied separately. For the transversal components the
$\zeta$--function reads 
\begin{equation}\label{h2}
\zeta^{T}(z)=\sum_{n=0}^{\infty}\sum_{m=1}^{\infty} D^{T}_{n,m}
(\lambda^{T}_{n,m})^{-z}+
\sum_{n=1}^{\infty}D^{T}_{n,0}
(\lambda^{T}_{n,0})^{-z}~~~,
\end{equation}
where the degeneracies are given by (\ref{v19}), (\ref{v27}),
while, according to Eq.(\ref{h1}), the eigen--values 
$\lambda^{T}_{n,m}$ are
\begin{equation}\label{h3}
\lambda^{T}_{n,m}=
\Lambda^{T}_{n,m}+a^{-2}(d-1)=
{1 \over a^2}\left(
(n+\gamma m)
(n+\gamma m +d-1)+d-2\right)~~~.
\end{equation}
The $\zeta$--function is known to be the key object of the one-loop
quantum--field computations on curved backgrounds \cite{BirDev}. In
the present paper we use this function to study the coefficients in
the Schwinger--DeWitt expansion of the heat kernel operator $K^T(s)$ 
\begin{equation}\label{h4}
K^T(s)=
\sum_{n=0}^{\infty}\sum_{m=1}^{\infty} D^{T}_{n,m}
e^{-s\lambda^T_{n,m}}+
\sum_{n=1}^{\infty}D^{T}_{n,0}
e^{-s\lambda^T_{n,0}}~~~.
\end{equation}
The quantities $\zeta^T(z)$ and $K^T(s)$ are related via the Mellin
transformation\footnote{In the presence of zero modes, which is not
our case, the relation (\ref{h5}) modifies.} 
\begin{equation}\label{h5}
\zeta^T(z)={1 \over \Gamma(z)} \int_{0}^{\infty}
ds~ s^{1-z}K^T(s)~~~,
\end{equation}
where $\Gamma(z)$ is the gamma--function. It enables one to express
the coefficients $A_{n}$ in the asymptotic expansion of $K^T(s)$ at
$s\rightarrow 0$ 
\begin{equation}\label{h6}
K^T(s)={1 \over (4\pi s)^d}\sum_{n=0,1,2..}^{\infty}A^T_n s^n
\end{equation}
in terms of the residues of the $\zeta$--function \cite{Camporesi}
\begin{equation}\label{h7}
A^T_n=(4\pi)^{d/2}\Gamma\left(\frac d2-n\right)
\mbox{Res}\left[\zeta^{T}\left({d \over 2} - n \right)\right]~~~,
\end{equation}
for an arbitrary large dimension $d>2n$, and as
\begin{equation}\label{h8}
A^T_{d/2}=(4\pi)^{d/2}\zeta^T(0)
\end{equation}
if $d=2n$. We will be interested in the coefficients $A_1$ and
$A_2$ because they are important for studying the
ultraviolet divergences and conformal anomalies in four
dimensions.

The representation of the $\zeta$--function which is
convenient for using Eqs.(\ref{h7}) and (\ref{h8}) is found in 
Appendix B
\begin{equation}\label{b1}
\zeta^T(z)=\zeta_1(z)+\zeta_2(z)+\zeta_3(z)~~~,
\end{equation}

\bigskip

$$
\zeta_1(z)={2(d-1) \over \gamma (d-2)!}\sum_{k=0}^{\infty}
C_k(z)\sigma^k\sum_{p=0}^{d-2}\sum_{l=0}^{p}
\left(
\begin{array}{c}
p\\l\\
\end{array}
\right)
w_p\sum_{q=0}^{p-l}
\left(
\begin{array}{c}
p-l\\q\\
\end{array}
\right)q!
$$
\begin{equation}\label{b21}
\times(-1)^{p+l}\left({d-1 \over 2}\right)^{p-l-q}
{\Gamma(2z+2k-l-q-1) \over \Gamma(2z+2k-l)}\zeta_R\left(2z+2k-l-q-1,
{d-1 \over 2}\right)~~~,
\end{equation}

\bigskip

$$
\zeta_2(z)={2(d-1) \over (d-2)!}\sum_{k=0}^{\infty}
C_k(z)\sigma^k\sum_{r=0}^{\infty}\sum_{p=0}^{d-2}\sum_{l=0}^{p}
\left(
\begin{array}{c}
p\\l\\
\end{array}
\right)
w_p\sum_{q=0}^{p-l}
\left(
\begin{array}{c}
p-l\\q\\
\end{array}
\right)(-1)^{p+l+q}\left({d-1 \over 2}\right)^{p-l-q}
$$
\begin{equation}\label{b22}
\times {B_{r+q+1} \over r+q+1} {\gamma^{r+q} \over r!}
{\Gamma(2z+2k+r-l) \over \Gamma(2z+2k-l)}\zeta_R\left(2z+2k+r-l,
{d-1 \over 2}\right)~~~,
\end{equation}

\bigskip

$$
\zeta_3(z)={1 \over (d-2)!}\sum_{k=0}^{\infty}
C_k(z)\sigma^k\sum_{p=0}^{d-3}\sum_{l=0}^{p}
\left(
\begin{array}{c}
p\\l\\
\end{array}
\right)
w'_p(-1)^{p+l}\left({d+1 \over 2}\right)^{p-l}
$$
\begin{equation}\label{b23}
\times \left[\frac 12 (d-3)^2\zeta_R\left(2z+2k-l,
{d+1 \over 2}\right)+(d-1)\zeta_R\left(2z+2k-l-1,
{d+1 \over 2}\right) \right] ~~~.
\end{equation}
In these expressions
\begin{equation}\label{h9}
\sigma={(d-3)^2 \over 4}~~~,
\end{equation}
\begin{equation}\label{h10}
C_k(z)={\Gamma(z+k) \over k! \Gamma(z)}~~~,
\end{equation}
$B_q$ are the Bernoulli numbers and quantities $w_p$, $w_p'$ are
defined as follows:
\begin{equation}\label{h11}
\sum_{p=0}^{d-2}n^p w_p(d)={(d+n-2)! \over n!}~~~,
\end{equation}
\begin{equation}\label{h12}
\sum_{p=0}^{d-3}n^p w'_p(d)={(d+n-2)! \over (n+2)n!}~~~.
\end{equation}
In the functions $\zeta_i(z)$ we have put for simplicity $a=1$,
because the dependence on the radius $a$ factorizes in a overall
multiplier $a^{2z}$, and thus it can be restored in the end of
calculation. The expressions (\ref{b21})--(\ref{b23}) hold when
$z\rightarrow 0,-1,-2,...$ . Although these formulas look complicated,
they are very convenient for our purpose. Indeed, under more close
examination one can see that for $\zeta_i(0)$ the series reduce to a
finite sum, while in the limit $z\rightarrow -1,-2,...$ the functions
$\zeta_i(z)$ have only a finite number of poles. In principle
Eqs.(\ref{b21})--(\ref{b23}) enable one to find the heat-kernel
coefficients $A_n$ with the help of Eq.(\ref{h7}) for arbitrary $d$
and $n$. 

In particular, with the help of Eqs.(\ref{z91})--(\ref{z10})
and after tedious but simple computations one obtains
\begin{equation}\label{h14}
{\bar{A}_1^{T}(d) \over (4\pi)^{d/2}}=
{\Gamma\left(\frac d2-1\right) \over 12 \gamma (d-3)!}
\left[d(d-7)+(d-1)\left(\gamma^2 -1\right) +
12\left(1-\gamma\right)\right]+\delta_{d,2}~~~,
\end{equation}
$$
{\bar{A}_2^T(d) \over (4\pi)^{d/2}}=
{\Gamma\left(\frac d2\right) \over 360 \gamma (d-1)!}
\left( 5d^5-77d^4+325d^3-369d^2+246d+(d-1)(d-3)\right.
$$
\begin{equation}\label{h15}
\left.
\times \left[2(d-1)
(1-\gamma^4) + 10 (d-1)(d-4)(\gamma^2-1)
+120(d-2)(1-\gamma)\right]\right)+\delta_{d,4}~~~,
\end{equation}
where $\delta_{d,k}$ is the Kronecker symbol. The bar on the above
coefficients stresses that they are computed on the singular spaces.
The origin of the last terms in the right hand side (r.h.s.) of
Eqs.(\ref{h14}) and (\ref{h15}) is explained by the different
expressions of the heat coefficients when $d>2n$ and $d=2n$, see Eqs.
(\ref{h7}), (\ref{h8}). Note that the results (\ref{h7}), (\ref{h8})
are obtained for $\beta \leq 2 \pi$, and they can be analytically
continued to $\beta > 2 \pi$. 

\subsection{$A_1$--coefficient} 

Consider now the complete coefficients $A_n^{(1)} $ of the vector  
Hodge-deRham operator. On $S^d_\beta$ they are the sum 
\begin{equation}\label{h16}
\bar{A}_n^{(1)}=\bar{A}_n^{T}+\bar{A}_n^{L}=\bar{A}_n^{T}+\bar{A}_n^{(0)}-
(4\pi)^{d/2} \delta_{d,2n}
\end{equation}
of the heat coefficients $\bar{A}_n^{T}$ and $\bar{A}_n^{L}$
corresponding to the transversal and longitudinal components,
respectively. On the other hand, the longitudinal coefficient
$\bar{A}_n^{L}$ is related to the coefficient $\bar{A}_n^{(0)}$ of the
scalar Laplacian from which one has to remove the contribution of the
single zero mode when $n=d/2$, see Eq.(\ref{v1}). 

For the scalar Laplacian on a closed 
space ${\cal M}_\beta$ with conical singularities having deficit
angle $2\pi-\beta$ and forming the hypersurface $\Sigma$
\begin{equation}\label{h17}
\bar{A}_1^{(0)}=\frac 16 \int _{{\cal M}_\beta -\Sigma}R +
A_{\beta,1}^{(0)}~~~, 
\end{equation}
\begin{equation}\label{h17b}
A_{\beta,1}^{(0)}={\pi \over 3\gamma}(\gamma^2-1)\int_{\Sigma}~~~,
\end{equation}
where, as before, $\gamma=2\pi/\beta$.
The general expression of the vector coefficient is \cite{FM96}
\begin{equation}\label{h18}
\bar{A}_1^{(1)}={ d-6 \over 6}\int _{{\cal M}_\beta -\Sigma}R +
A_{\beta,1}^{(1)}~~~,
\end{equation}
\begin{equation}\label{h18b}
A_{\beta,1}^{(1)}=d~A_{\beta,1}^{(0)}
+{4\pi \over \gamma}(1 -\gamma)\int_{\Sigma}~~~.
\end{equation}
Note that $A_{\beta,1}^{(0)}$, $A_{\beta,1}^{(1)}$ are the corrections
to the heat coefficients due to conical singularities. These
corrections are proportional to the volume of the surface $\Sigma$,
which is $S^{d-2}$ in the case under consideration. One can
check\footnote{ It is worth reminding that the volume of $S^d_\beta$
is $(4\pi)^{d/2}\Gamma(d/2)/(\gamma (d-1)!)$.} by substitution
of (\ref{h14}) and (\ref{h17}) into
Eq.(\ref{h16}) with $n=1$, that it reproduces the exact formula 
(\ref{h18}) for $d$--spheres $S^d_\beta$. Therefore, the direct 
mode-by-mode computation
of the first vector coefficient confirms the general analysis of
\cite{FM96}. 

\subsection{$A_2$--coefficient} 
The form of this coefficient in vector case is not yet known
on singular spaces. On the smooth closed manifolds
(see for instance \cite{DeWitt,BV})
\begin{equation}\label{h19}
A_2^{(0)}={1 \over 180}\int_{{\cal M}}
\left(R_{\mu\nu\lambda\rho}R^{\mu\nu\lambda\rho}-R_{\mu\nu}R^{\mu\nu}
+\frac 52 R^2\right)~~~,
\end{equation}
\begin{equation}\label{h20}
A_2^{(1)}=d~A_2^{(0)}+
\int_{{\cal M}}\left(-{1 \over 12}R_{\mu\nu\lambda\rho}R^{\mu\nu\lambda\rho}
+\frac 12 R_{\mu\nu}R^{\mu\nu}-\frac 16 R^2\right)~~~,
\end{equation}
where $R_{\mu\nu\lambda\rho}$ and  $R_{\mu\nu}$ are the components of
the Riemann and Ricci tensors respectively. First of all, with the
help of Eqs.(\ref{h16}),(\ref{h19}) and (\ref{h20}) one can check that
$A_2^T$, Eq.(\ref{h15}), has the correct expression on $S^d$ when
$\gamma=1$. 

Another interesting exercise is the comparison of (\ref{h15}) at the
small deficit angle ($\gamma\simeq 1$) with the behaviour of
$A_2^{(1)}$ on manifolds with the "blunted" singularities \cite{FM96}.
Indeed, one can consider the coefficients (\ref{h19}) and (\ref{h20})
on a sequence of smooth manifolds $\tilde{\cal M}_\beta$ which
converges to a space ${\cal M}_\beta$ with conical singularities.
We refer to $\tilde{\cal M}_\beta $ as to manifolds with the blunted
singularities. One can prove \cite{FS95} that the following formulas: 
\begin{equation}\label{h22}
\int_{\tilde{\cal M}_\beta} R^2\simeq\int_{{\cal M}_\beta-\Sigma} R^2
+8\pi(\gamma-1)\int_{\Sigma} R~~~,
\end{equation}
\begin{equation}\label{h23}
\int_{\tilde{\cal M}_\beta} R_{\mu\nu}R^{\mu\nu} \simeq\int_{{\cal M}_\beta-
\Sigma} R_{\mu\nu}R^{\mu\nu}
+4\pi(\gamma-1)\int_{\Sigma} R_{ii}~~~,
\end{equation}
\begin{equation}\label{h24}
\int_{\tilde{\cal M}_\beta} R_{\mu\nu\lambda\rho}R^{\mu\nu\lambda\rho} 
\simeq\int_{{\cal M}_\beta-
\Sigma} R_{\mu\nu\lambda\rho}R^{\mu\nu\lambda\rho}
+8\pi(\gamma-1)\int_{\Sigma} R_{ijij}~~~,
\end{equation}
are valid up to the leading order in $(\gamma-1)$ when $\tilde{\cal
M}_\beta\rightarrow {\cal M}_\beta$ . Here $R_{ii}=
R_{\mu\nu}n^{\mu}_in^{\nu}_i$ and
$R_{ijij}=R_{\mu\nu\lambda\rho}n^{\mu}_i
n^{\nu}_jn^{\lambda}_in^{\rho}_j$ are the components of the Ricci and
Riemann tensors computed in the regular domain near $\Sigma$ and
normal to this surface. The quantities $n^{\mu}_i$ ($i=1,2$) are two
unit normal vectors. On the $d$--spheres with unit radius 
\begin{equation}\label{h25}
R=d(d-1)~~~,
\end{equation}
\begin{equation}\label{h26}
R_{ii}=2(d-1)~~~,
\end{equation}
\begin{equation}\label{h27}
R_{ijij}=2~~~.
\end{equation}
Thus, by taking into account Eqs.(\ref{h16}), (\ref{h19})--(\ref{h27})
one can verify that for small values of $|1-\gamma|$ the coefficients
(\ref{h15}) can be approximated by the standard coefficients evaluated
on the spheres $\tilde{S}^d_\beta$ with the blunted conical
singularities 
\begin{equation}\label{h28}
\bar{A}_2^T[S^d_\beta]=A_2^{(1)}[\tilde{S}^d_\beta]-
A_2^{(0)}[\tilde{S}^d_\beta]+(4\pi)^{d/2}\delta_{d,2n}
+O((1-\gamma)^2)~~~. 
\end{equation}
As it was shown in \cite{FS96}, the same property holds in general for
the coefficients of the scalar Laplacians (without curvature
couplings). Hence, one can also conclude that 
\begin{equation}\label{h29}
\bar{A}_2^{(1)}[S^d_\beta]=A_2^{(1)}[\tilde{S}^d_\beta]
+O((1-\gamma)^2)~~~. 
\end{equation}
Note that Eq.(\ref{h29}) is a nice property of the vector Hodge-deRham
operator $\triangle^{(1)}$ only. Arbitrary vector Laplacians, like
$-\nabla_\mu \nabla^\mu$ for example, which differ from
$\triangle^{(1)}$ by terms depending on curvature, do not have this
property. 
 
\bigskip

Let us discuss now the form of $\bar{A}^{(1)}_2$ on arbitrary manifolds 
${\cal M}_\beta$ with conical singularities. Expression (\ref{h20})
for the standard coefficient suggests that this form can be similar
to that of the first vector coefficient, Eqs.(\ref{h18}),(\ref{h18b}),
\begin{equation}\label{h30}
\bar{A}^{(1)}_2=A^{(1)}_2+A^{(1)}_{\beta,2}~~~,
\end{equation}
\begin{equation}\label{h31}
A^{(1)}_{\beta,2}=d ~A^{(0)}_{\beta,2}+ Q_{\beta,2}~~~.
\end{equation}  
Here $A^{(0)}_{\beta,2}$ is the correction to the scalar coefficient
$A^{(0)}_2$ (\ref{h19}) due to conical singularities
\cite{F95b}
\begin{equation}\label{h32}
A^{(0)}_{\beta,2}={\pi \over 90\gamma}\int_{\Sigma}\left[
(1-\gamma^4)\left(\frac 12R_{ii}-R_{ijij}\right)-5(1-\gamma^2)R\right]
~~~.
\end{equation}
$Q_{\beta,2}$ is a functional similar to $A^{(0)}_{\beta,2}$
$$
Q_{\beta,2}={\pi \over \gamma}\int_{\Sigma}\left[(1-\gamma^4)
(a_1R+a_2 R_{ii}+a_3 R_{ijij})\right.
$$
\begin{equation}\label{h33}
\left.
+
(1-\gamma^2)(b_1R+b_2 R_{ii}+b_3 R_{ijij})+
(1-\gamma)(c_1R+c_2 R_{ii}+c_3 R_{ijij})\right]~~~,
\end{equation}
where $a_k$, $b_k$ and $c_k$ are numerical coefficients. This
structure of $Q_{\beta,2}$ results from considerations similar to
those of Dowker \cite{Dowker94}. Let the dimensionality of the
parameter $s$ in the Schwinger-DeWitt expansion be $L^2$, where $L$ is
a length. Then, as it follows from (\ref{h6}), the dimensionality of
$Q_{\beta,2}$ is $L^{d-4}$. On the other hand, the integral
$\int_{\Sigma}$ scales as $L^{d-2}$ and, consequently the integrand in
(\ref{h33}) has the dimension $L^{-2}$. According to the Gauss-Codacci
equations, see Ref.\cite{F95b}, one can find only three independent
invariants with such dimension: $R$, $R_{ii}$ and $R_{ijij}$, which
describe internal and external geometry of $\Sigma$. 

Fortunately, all the coefficients in (\ref{h33}) can be fixed if one
makes use of Eq.(\ref{h16}) and expression (\ref{h15}) for the
transversal vectors on $S^d_\beta$. After simple computations we get 
Eq. (\ref{i31})
\begin{equation}\label{h34}
A^{(1)}_{\beta,2}=d~A^{(0)}_{\beta,2}+{\pi \over 3\gamma}
\int_{\Sigma}\left[(1-\gamma^2)(R-R_{ii})+2(1-\gamma)
(R-2 R_{ii}+R_{ijij})\right]~~~.
\end{equation}
It would be interesting to confirm this result by the 
direct computations analogous to those of \cite{F95b},\cite{Dowker:94PR}
without assumptions (\ref{h32}),(\ref{h33}). 

\section{Renormalization of the gauge effective action}
\setcounter{equation}0

Let us discuss now renormalization in quantum field theory on singular
spaces  ${\cal M}_\beta$. As it was mentioned in the Section 1 the
geometrical structure of the divergences $W_{\mbox{div}}$ in quantum
theory on curved backgrounds is determined by the coefficients $A_0$,
$A_1$ and $A_2$, see (\ref{i12}). The effective action $W$ for the
gauge fields quantized with the gauge condition $\nabla^\mu V_\mu=0$
is given in terms of the determinant of the Hodge-deRham operator by
Eq. (\ref{i11}), while the contribution of ghosts has the form
of the determinant of the scalar Laplacian. 
It follows from the results of Ref.\cite{FM96} and
Eq.(\ref{h34}) for $A_{\beta,2}^{(1)}$,  that in the leading
order in deficit angle $(2\pi-\beta)$ 
\begin{equation}\label{g1}
\bar{A}_{n}^{(1)}[{\cal M}_\beta]\simeq A_{n}^{(1)}[\tilde{\cal M}_\beta]
+O((2\pi-\beta)^2)~~~,~~~n=1,2~~~.
\end{equation}
It means that one can approximate the coefficients on the singular
spaces ${\cal M}_\beta$ by their expressions on the
corresponding smooth manifolds $\tilde{\cal M}_\beta$.
As for $A_0$, it does not depend on conical singularities.
Then Eq.(\ref{i11}) says that the same property is valid for the
divergent part of the gauge action
\begin{equation}\label{g2}
W_{\mbox{div}}[{\cal M}_\beta]\simeq W_{\mbox{div}}[\tilde{\cal M}_\beta]
+O((2\pi-\beta)^2)~~~.
\end{equation}
Hence, in this order the one-loop divergences on ${\cal M}_\beta$ are
completely removed by the {\it standard} renormalization 
procedure \cite{BirDev} used on the smooth backgrounds, i.e. by the
renormalization of the Newton constant and couplings by the terms
$R^2$, $R_{\mu\nu}^2$ and $R_{\mu\nu\lambda\rho}^2$ in the
bare gravitational Lagrangian. The
effective action of the scalar fields without curvature couplings has
the analogous property \cite{FS96} and, hence, so does the ghost 
action.

This conclusion is important for black hole thermodynamics in the
off-shell formulation \cite{BTZ}-\cite{MSb}. In this case ${\cal
M}_\beta$ appear as the Euclidean section of the corresponding
Lorentzian space-times, $\Sigma$ represents the Euclidean horizon and 
the imaginary time period $\beta$ is associated with the inverse
temperature of the system. The regularity condition $\beta=2\pi$
corresponds to the Hawking temperature. For off-shell computations of
the black hole entropy $S$ it is sufficient to
use the decomposition of the effective
action $W$ near $\beta=2\pi$ up to the terms $\sim (2\pi-\beta)$. Our
results mean that in case of gauge fields the ultraviolet divergences
in $S$ are removed under the standard renormalization of coupling
constants in the bare gravitational action, analogously to scalar
fields \cite{FS96}. It also indicates the agreement between on-shell and
off-shell computations of the thermodynamical quantities 
for black holes \cite{FFZ}. 

\section{Summary and perspectives}
\setcounter{equation}0

The main result of this paper is the exact computation of the spectrum
(\ref{d1})--(\ref{deg2}) of the vector Laplacian on $d$--spheres
$S^d_\beta$ with the conical singularities. This is the generalization
of the results obtained by Rubin and Ordonez \cite{Ordonez} on usual
spheres $S^d$. The spectrum can be used now for the computation of the
gauge one-loop effective action in terms of the $\zeta$--function
(\ref{b1})--(\ref{b23}). This will be a development of works by Allen
et al. \cite{Allen}-\cite{emr} investigated the effective potential in
gauge models in the de Sitter space. 

In the present paper we used the spectrum  (\ref{d1})--(\ref{deg2}) to
compute the coefficients of the heat kernel expansion on $S^d_\beta$
and to study their properties. Computation of $\bar{A}^{(1)}_1$
confirms the results of \cite{Kabat} and \cite{FM96} found by
different methods. We have also calculated the second coefficient
$\bar{A}^{(1)}_2$ on $S^d_\beta$ and suggested its generalization
(\ref{h34}) to arbitrary manifolds ${\cal M}_\beta$ with conical
singularities. This result is both new and important because
$\bar{A}^{(1)}_2$ is related to the conformal anomaly. 

Finally we have shown that the vector Hodge-deRham operator on ${\cal
M}_\beta$ has a number of properties analogous to those of the scalar
Laplacian $-\nabla^\mu\nabla_\mu$. The heat coefficients for small
deficit angles can be approximated by the
coefficients on the corresponding spaces with the blunted
singularities. The immediate consequence of these properties is the
validity for gauge fields of a renormalization theorem concerning the
black hole entropy proven in \cite{FS96} for scalar fields. 

The method employed here to find the spectrum (\ref{d1})--(\ref{deg2})
is straightforward but it is quite simple. It would be interesting to
use it to get the spectra of the Laplacians of rank 2 symmetric
tensors on $S^d_\beta$ and to continue investigation of these
operators on singular spaces started in \cite{FM96}. Studying this
problem is in progress. 

\noindent
{\bf Acknowledgements} One of the authors (D.V.F) is very grateful to
University of Naples for hospitality and to INFN for the support
during his visit to Naples. This work was also supported in part by
the Natural Sciences and Engineering Research Council of Canada. 
\newpage
\appendix

\section{Explicit solution of constraints}
\setcounter{equation}0

We present here the explicit 
solution of constraints (\ref{v13})--(\ref{v15}) and 
(\ref{cond21})--(\ref{cond24}) given in terms of 
the coefficients $(B^k)^m_{i_1,...,i_n}$ with $m=0,1,2,..$,
and 
$(B^{-})^0_{i_1,...,i_{n-1}}$. 
The method of its construction is described in Section 3.

If $m \neq 0$ the solution is
\begin{eqnarray}
(B^-)^m_{i_1,...,i_{n-2p+1}}&=&{(-1)^{p+1} \over 2^p p!}{n! \over (n-2p+1)!}
{\Gamma (\gamma m) \over \Gamma(\gamma m+p)}~~
(B_{\{ h_1})^m_{ h_1,...,h_p, h_p,i_1,...,i_{n-2p+1}\}}\nonumber\\
&~&~~~~~~~~~~~~~~~~~~~~~~~~~~~~~~~~~~~~~~\mbox{for}
~~~~0\leq p\leq \left[ {n+1 \over 2}\right]~~~,\label{v16} \\
(B^k)^m_{i_1,...,i_{n-2p}}&=&{(-1)^p \over 2^p p!}{n! \over (n-2p)!}
{\Gamma(\gamma m+1) \over \Gamma(\gamma m+p+1)!}~~
(B^k)^m_{h_1, h_1,...,h_p ,h_p,i_1,...,i_{n-2p}}\nonumber\\
&~&~~~~~~~~~~~~~~~~~~~~~~~~~~~~~~~~~~~~~~~~~~~~~\mbox{for}
~~~~1\leq p\leq \left[ {n \over 2}\right]~~~,\label{v17}
\end{eqnarray}
$$ 
(B^+)^m_{i_1,...,i_{n-2p-1}} = {(-1)^p \over 2^{p+1} p!}{n! \over (n-2p-1)!}
{\Gamma(\gamma m+2) \over \Gamma(\gamma m+p+2)}
$$
$$
\times\left[{1 \over \gamma m+1}
(B_{\{{h_1}})^m_{ h_1,...,h_p,h_p,i_1,...,i_{n-2p-1}\},h,h}
- {1 \over \gamma m} (B_{\{{h_1}})^m_{ h_1,...,h_{p+1}, h_{p+1},
i_1,...,i_{n-2p-1}\}} 
\right]~~~
$$
\begin{equation}
~~~~~~~~~~~~~~~~~~~~~~~~~~~~~~~~~~~~~~~~~~~~~~~~~~~~~~
\mbox{for}~~~0\leq p\leq \left[ {n-1 \over 2}\right]~~~.
\label{v18}
\end{equation} 

For $m=0$ one can find 
\begin{eqnarray}
(B^-)^0_{i_1,...,i_{n-2p+1}}&=&{(-1)^{p+1} \over 2^{(p-1)} p!(p-1)!}{(n-1)! 
\over (n-2p+1)!}
(B^-)^0_{h_1,h_1,...,h_{p-1},h_{p-1},i_1,...,i_{n-2p+1}}~~ \nonumber\\
&~&~~~~~~~~~~~~~~~~~~\mbox{for}
~~~~~2\leq p\leq \left[ {n+1 \over 2}\right]~~~,\label{v20}\\
(B^+)^0_{i_1,...,i_{n-2p-1}}&=&
{(-1)^{p+1} \over 2^p p!(p+1)!}{(n-1)! \over (n-2p-1)!}
\left[ (B^-)^0_{h_1,h_1,...,h_p,h_p,i_1,...,i_{n-2p-1}} \right.\nonumber\\
&+&\left.
n (B_{h_1})^0_{ h_1,...,h_{p+1},h_{p+1},i_1,...,i_{n-2p-1}}\right] 
~~~~\mbox{for}~~~~0\leq p \leq \left[ {n-1 \over 2}\right]~~~,
\label{v22}\\ 
(B^k)^0_{i_1,...,i_{n-2p}}&=&{(-1)^{p} \over 2^p(p!)^2}{n! \over (n-2p)!}
(B^k)^0_{h_1,h_1,...,h_p,h_p,i_1,...,i_{n-2p}}~~\mbox{for}~~~
0\leq p\leq \left[ {n\over 2}\right].\label{v23}
\end{eqnarray}
One can verify by a straightforward substitution that the solutions
(\ref{v16})--(\ref{v18}) and (\ref{v20})--(\ref{v23}) satisfy all
constraints (\ref{cond21})--(\ref{cond23}), (\ref{cond24}) and
relations (\ref{v13})--(\ref{v15}).

\section{$\zeta$--function}
\setcounter{equation}0

Here we give the details how to 
derive representation (\ref{b1})--(\ref{b23}) for the $\zeta$--function
We follow the method developed in \cite{FM}
for the scalar $\zeta$--function.

First, we use the decomposition
\begin{equation}\label{z1}
(\lambda^T_{n,m})^{-z}=\sum_{k=0}^{\infty}C_k(z)\sigma^k
\left(n+\gamma m +{d-1 \over 2}\right)^{-2z-2k}~~~,
\end{equation}
where $\sigma$ and $C_k(z)$ are defined by Eqs.(\ref{h9}),(\ref{h10})
respectively. Then, by taking into account definitions
(\ref{h11}) and (\ref{h12}) we obtain
$$
\zeta_1(z)+\zeta_2(z)=
\sum_{n=0}^{\infty}\sum_{m=1}^{\infty}D_{n,m}^T
(\lambda^T_{n,m})^{-z}
$$
\begin{equation}\label{z2}
={2(d-1) \over (d-2)!}\sum_{k=0}^{\infty}C_k(z)\sigma^k
\sum_{n=0}^{\infty}\sum_{m=1}^{\infty}
\left(n+\gamma m +{d-1 \over 2}\right)^{-2z-2k}\sum_{p=0}^{d-2}
n^pw_p~~~,
\end{equation}

\bigskip

$$
\zeta_3(z)=\sum_{n=1}^{\infty} D_{n,0}^T(\lambda^T_{n,0})^{-z}
$$
\begin{equation}\label{z3}
={1 \over (d-2)!}\sum_{k=0}^{\infty}C_k(z)\sigma^k
\sum_{n=0}^{\infty}
\left(n +{d+1 \over 2}\right)^{-2z-2k}
\left[d^2+d(n-3)+4-n\right]
\sum_{p=0}^{d-3}
n^pw'_p~~~,
\end{equation}
where in the r.h.s. of (\ref{z3}) the summation index $n$ was replaced
with $n+1$. Let us focus now on the computation of
$\zeta_1(z)+\zeta_2(z)$, because the computation of $\zeta_3(z)$ is analogous.
One can use in (\ref{z2}) the decomposition 
\begin{equation}\label{z4}
n^p=\sum_{l=0}^p(-1)^{p-l}
\left(
\begin{array}{c}
p\\l\\
\end{array}
\right)
\left(n+\gamma m +{d-1 \over 2}\right)^l
\left(\gamma m +{d-1 \over 2}\right)^{p-l}
\end{equation}
to get
$$
\zeta_1(z)+\zeta_2(z)=
{2(d-1) \over (d-2)!}\sum_{k=0}^{\infty}C_k(z)\sigma^k
$$
\begin{equation}\label{z5}
\times
\sum_{p=0}^{d-2}\sum_{l=0}^{p}(-1)^{p+l}
\left(
\begin{array}{c}
p\\l\\
\end{array}
\right)
w_p
\sum_{m=1}^{\infty}
\zeta _R\left(2z+2k-l,\gamma m+{d-1 \over 2}\right)
\left(\gamma m +{d-1 \over 2}\right)^{p-l}~~~,
\end{equation}
where
$$
\zeta _R(z,a)=\sum_{n=0}^{\infty}(n+a)^{-z}~~~,~~~a\neq 0,-1,-2,...
$$
is the Riemann $\zeta$--function. The integral representation 
of $\zeta_R(z,a)$
\begin{equation}\label{z6}
\zeta_R (z,a)={1 \over \Gamma(z)}\int_{0}^{\infty}
{y^{z-1}e^{-ay} \over 1-e^{-y}}dy
\end{equation}
enables one to sum up over $m$ in Eq.(\ref{z5})
$$
\sum_{m=1}^{\infty}
\zeta _R\left(2z+2k-l,\gamma m+{d-1 \over 2}\right)
\left(\gamma m +{d-1 \over 2}\right)^{p-l}
$$
\begin{equation}\label{z7}
=
{1 \over \Gamma(2z+2k-2l)}
\int_{0}^{\infty} dy~ {y^{2z+2k-l-1} \over 1-e^{-y}}
(-1)^{p-l}\left({d \over dy}\right)^{p-l}
\left[{\exp\left(-{d-1 \over 2}y\right) \over e^{\gamma y} -1}
\right]~~~.
\end{equation}
Then one can use in (\ref{z7}) 
the definition of the Bernoulli numbers $B_n$
$$
{x \over e^x-1}=\sum_{n=0}^{\infty} {B_n \over n!} x^n~~~.
$$
This gives
$$
\sum_{m=1}^{\infty}
\zeta _R\left(2z+2k-l,\gamma m+{d-1 \over 2}\right)
\left(\gamma m +{d-1 \over 2}\right)^{p-l}
$$
$$
={1 \over \gamma}\sum_{q=0}^{p-l}
\left(
\begin{array}{c}
p-l\\q\\
\end{array}
\right)
\left({d-1 \over 2}\right)^{p-l-q}q!
{\Gamma(2z+2k-l-q-1) \over \Gamma(2z+2k-l)}
\zeta_R\left(2z+2k-l-q-1,{d-1 \over 2}\right)
$$
$$
+
\sum_{r=0}^{\infty}
\sum_{q=0}^{p-l}
\left(
\begin{array}{c}
p-l\\q\\
\end{array}
\right)
\left({d-1 \over 2}\right)^{p-l-q}
(-1)^q
{B_{r+q+1} \over r+q+1}
{\gamma ^{r+q} \over r!}
$$
\begin{equation}\label{z8}
\times {\Gamma(2z+2k+r-l) \over \Gamma(2z+2k-l)}
\zeta_R\left(2z+2k+r-l,{d-1 \over 2}\right)~~~.
\end{equation}
Now the substitution of (\ref{z8}) into (\ref{z5}) gives the 
expressions for $\zeta_1(z)$ and $\zeta_2(z)$. The first term
in r.h.s. of (\ref{z8}) results in expression (\ref{b21}), while 
the second term in r.h.s. of (\ref{z8}) 
corresponds to formula (\ref{b22}).

Finally we give explicitly several numbers $w_p$, see 
Eq.(\ref{h11}), which are used to get Eqs.(\ref{h14}) and 
(\ref{h15})
\begin{equation}\label{z91}
w_{d-2}=1~~~,
\end{equation}
\begin{equation}\label{z92}
w_{d-3}=\frac 12(d-1)(d-2)~~~,
\end{equation}
\begin{equation}\label{z93}
w_{d-4}={1 \over 24}(d-1)(d-2)(d-3)(3d-4)~~~,
\end{equation}
\begin{equation}\label{z94}
w_{d-5}={1 \over 48}(d-1)^2(d-2)^2(d-3)(d-4)~~~,
\end{equation}
\begin{equation}\label{z95}
w_{d-6}={1 \over 5760}(d-1)(d-2)(d-3)(d-4)(d-5)(15d^3-75d^2+110d-48)~~~.
\end{equation}
Regarding the numbers $w_p'$ in formula (\ref{h12}), they can be found 
from $w_p$ with the help of relations
\begin{equation}\label{z10}
w_{d-3}'=1~~~,~~~w_{p-1}'=w_p-2w_p'~~~,
\end{equation}
for $1\leq p \leq d-4$.

\end{document}